\documentclass{easychair}

\usepackage{float}
\usepackage{amsmath}
\usepackage{graphicx}
\usepackage{semantic}
\usepackage{xspace}
\usepackage{pgffor}
\usepackage{tikz}
\usepackage{hyperref}
\usetikzlibrary{fit}

\begin{document}

\title{On SMT Theory Design: The Case of Sequences}

\author{Hichem Rami Ait El Hara\inst{1, 2}
  \and
  François Bobot\inst{2}
  \and
  Guillaume Bury\inst{1}
}

\titlerunning{On SMT Theory Design: The Case of Sequences}

\authorrunning{Ait El Hara et al.}

\institute{OCamlPro, Paris, France \and
  Université Paris-Saclay, CEA, List, F-91120, Palaiseau, France}

\maketitle

\begin{abstract}

  Choices in the semantics and the signature of a theory are integral in determining how the theory is used and how challenging it is to reason over it. Our interest in this paper lies in the SMT theory of sequences. Various versions of it exist in the literature and in state-of-the-art SMT solvers, but it has not yet been standardized in the SMT-LIB. We reflect on its existing variants, and we define a set of theory design criteria to help determine what makes one variant of a theory better than another. The criteria we define can be used to appraise theory proposals for other theories as well. Based on these criteria, we propose a series of changes to the SMT theory of sequences as a contribution to the discussion regarding its standardization.

\end{abstract}

\newcommand{\ite}{\textsf{ite}\xspace}
\newcommand{\llet}{\textsf{let}\xspace}
\newcommand{\Array}{\textsf{Array}\xspace}
\newcommand{\Seq}{\textsf{Seq}\xspace}
\newcommand{\Seqc}{$\textsf{Seq}_{\textsf{cvc5}}$\xspace}
\newcommand{\Seqz}{$\textsf{Seq}_{\textsf{z3}}$\xspace}
\newcommand{\Arrayc}{$\textsf{Array}_{\textsf{c}}$\xspace}

\newcommand{\Tsa}{\textsf{Seq}(\alpha)\xspace}
\newcommand{\Elem}{\textsf{Elem}\xspace}
\newcommand{\Tint}{\textsf{Int}\xspace}
\newcommand{\Tbool}{\textsf{Bool}\xspace}
\newcommand{\Tt}{\textsf{T}\xspace}

\newcommand{\MyNewOperatorAux}[2]{\expandafter\newcommand\csname #1\endcsname{\ensuremath{\textsf{#2}}\xspace}}
\newcommand{\MyNewOperator}[1]{\MyNewOperatorAux{#1}{#1}}
\newcommand{\NewSeqOperator}[1]{\MyNewOperatorAux{#1}{seq.#1}}

\MyNewOperatorAux{sempty}{seq.empty}
\NewSeqOperator{unit}
\MyNewOperatorAux{slength}{seq.len}
\NewSeqOperator{nth}
\NewSeqOperator{extract}
\MyNewOperatorAux{sconcat}{seq.concat}
\NewSeqOperator{at}
\NewSeqOperator{contains}
\NewSeqOperator{indexof}
\NewSeqOperator{replace}
\NewSeqOperator{prefixof}
\NewSeqOperator{suffixof}

\MyNewOperatorAux{replaceall}{seq.replace\_all}
\NewSeqOperator{rev}
\NewSeqOperator{update}

\NewSeqOperator{map}
\NewSeqOperator{mapi}
\MyNewOperatorAux{foldleft}{seq.fold\_left}
\MyNewOperatorAux{foldlefti}{seq.fold\_lefti}

\NewSeqOperator{slice}
\MyNewOperatorAux{lengths}{length}
\MyNewOperatorAux{nths}{nth}
\MyNewOperatorAux{repeats}{repeat}
\MyNewOperatorAux{apps}{app}
\MyNewOperatorAux{updates}{update}
\MyNewOperatorAux{slices}{slice}
\MyNewOperatorAux{mapf}{map}
\MyNewOperatorAux{sset}{seq.set}
\MyNewOperatorAux{sget}{seq.get}
\MyNewOperatorAux{sconst}{seq.const}

\MyNewOperator{select}
\MyNewOperator{store}

\section{Introduction}

The SMT theory of arrays, introduced by McCarthy \cite{mccarthy_towards_1962}, maps index terms of a given sort to value terms of another sort. The theory provides two functions: the $\select(a, i)$ function, which takes an array $a$ and index $i$, and returns the value stored at $i$ in $a$; and the $\store(a,i,v)$ function, which takes an array $a$, an index $i$, and a value $v$, and returns a modified copy of $a$ in which the value $v$ is at the index $i$. This theory is minimal and generic, and many efficient decision procedures have been proposed for it \cite{de_moura_generalized_2009, brummayer_lemmas_2008, christ_weakly_2015}. However, the theory’s lack of expressiveness hinders reasoning on more complex data structures, as it makes it necessary to define additional functions and to use axiomatization, eventually with quantifiers, to express properties on such data structures.

Thus, when it comes to verifying properties of a given data structure using SMT solvers, it is more convenient to have a tailored theory that clearly and concisely describes the semantics of the higher-level operations on that data structure. Not only does this make verification easier for the user, but it can also pave the way for more dedicated and efficient decision procedures for the theory. Examples of such theories are the theory of strings and the theory of sequences, which have both sparked a lot of interest in recent years.

Sequences are a common data structure in programming languages, although they may be known by different names and have various implementations. Sequences can have fixed sizes, like arrays in C, C++, Rust, OCaml, and Java, or they can be dynamic, like vectors in C++ and Rust, ArrayLists in Java, arrays in JavaScript, and lists in Python. In addition to the common array operations, such as storing and selecting values at an index, some languages support higher-level operations such as concatenation, slicing, mapping, filtering, and folding. Recent studies tend to confirm our hypothesis: it is more efficient to verify properties of such data structures with an SMT solver by using the theory of sequences rather than the theory of arrays with additional axiomatization \cite{sheng_reasoning_2023}.

The theory of sequences differs from the theory of arrays in that sequences are dynamic and can change in size, while arrays have a fixed size determined by the sort of the indices and the number of possible values it has. Sequences are always indexed by integers, whereas arrays don't have such restrictions on their index sort. Additionally, the signature of the theory of sequences is richer than that of the theory of arrays; it includes functions for concatenation, slicing, subsequence extraction, etc. The $\textsf{nth}(s,i)$ function from the theory of sequences takes a sequence $s$ and an index $i$, returning the value stored at the $i$th index of the sequence, akin to the $\select$ function in the array theory. However, the mathematical interpretation of this function on sequences is partial to valid indices, which are those within the bounds of the sequence. As the SMT-LIB is a total logic, such functions are totalized.

A function is considered partial when it is not defined for all possible inputs, thus being applicable outside its domain. The returned value in such cases depends on how the function is totalized. Partial function totalization can be achieved in three ways. Firstly, through underspecification, which consists in returning an uninterpreted value. An uninterpreted value has no associated interpretation, meaning it is unconstrained and can be any value of the right sort. Secondly, overspecification, which entails returning a default constant value when a partial function is applied outside its domain. Thirdly, the additional argument approach, where an argument is added to the function. This argument will be returned when the function is applied outside its domain, allowing the user to determine the return value by providing that value to the function.

In this paper, we aim to discuss the design of the theory of sequences. We will define a set of criteria that ought to be taken into consideration when designing an SMT theory. Additionally, we will describe the various approaches used in the literature and in SMT solvers to totalize partial functions. Then, we will apply the criteria we have defined to suggest some variations to the theory of sequences. These variations aim to improve the theory for both those trying to reason over it and the users of the theory.

The paper is organized as follows: We begin by stating the notation we use in Section 2. Section 3 summarizes the known theories of sequences in the literature and in SMT solvers, and discusses the differences between them. In Section 4, we define a set of SMT theory design criteria that we consider as important to take into account when designing an SMT theory. Section 5 explores how partial functions are dealt with in SMT literature and discusses which approach is adequate when. In Section 6, we propose variations of the theory of sequences that we have designed, taking into account the criteria described in Section 4, and discuss why we believe they are better. Finally, we conclude in Section 7.

\paragraph{Related work}

Regarding theory design, some standardized theories in the SMT-LIB are actually based on published work proposing signatures and semantics for these theories. t is the case for the theory of Floating Point arithmetic \cite{rummer_smt-lib_2010,brain_automatable_2015} and the theory of strings \cite{bjorner_smt-lib_2012}. To our knowledge, theory design is not discussed in other contributions as clearly as it is here, but there are extensive discussions on the subject available on the SMT-LIB mailing list\footnote{Available at: https://groups.google.com/g/smt-lib}.

On the subject of handling partial functions, it is a known problem in mathematical logic in general. There are few works on it in Satisfiability Modulo Theories \cite{berezin_practical_2005}. However, there is more significant literature on the topic in related fields such as HOL (Higher Order Logic) and proof assistant development \cite{schmitt_computer-assisted_2011,hahnle_many-valued_2005,goos_avoiding_1995}.

The theory of sequences was introduced relatively recently, it was first formalized by Bjørner et al. \cite{bjorner_smt-lib_2012}. Sheng et al. developed calculi to reason over their variant of the theory of sequences \cite{sheng_reasoning_2023}. Additionally, there are works on the theory of arrays that extend the theory with a length function \cite{bonacina_cdsat_2022, ghilardi_interpolation_2023}, a concatenation function \cite{wang_solver_2023}, and others \cite{bradley_whats_2006, SMT2012:Theory_of_Arrays_with, ge_complete_2009}, giving arrays similar properties to those of sequences.

\section{Notation}

\textsf{Array} is the sort of arrays, \Elem is the sort of values, \Tint is the sort of integers, \Seq is the sort of sequences, and \Tbool is the sort of boolean values. The symbols $\min$ and $\max$ denote the usual mathematical functions. $\ite$ is the commonly used function in the SMT-LIB Standard. The symbol $\alpha$ represents a sort variable and can be any sort. The sort of higher-order functions $(\Tt_1 -> ... -> \Tt_{n-1} -> \Tt_n)$ is used as a simplification for the sort $\Array \Tt_1 (\Array ... (\Array \Tt_{n-1} \Tt_n))$. In $\llet\ x = v, y$, the $\llet$ symbol binds a variable $x$ to a term $v$ in a term $y$.

\section{The theory of sequences}

Sequences are 0-indexed ordered collections of elements with dynamic lengths. The SMT theory of sequences was proposed by Bjørner et al. \cite{bjorner_smt-lib_2012} as a generalization of the theory of strings to non-character values (\Seq). State-of-the-art SMT solvers such as CVC5\footnote{CVC5's sequence theory: \url{https://cvc5.github.io/docs-ci/docs-main/theories/sequences.html}} \cite{sheng_reasoning_2023} and Z3\footnote{Z3's sequence theory: \url{https://microsoft.github.io/z3guide/docs/theories/Sequences}} support theories of sequences, referred to as \Seqc and \Seqz respectively. Their signatures share many symbols with \Seq, along with some additions and deductions.

Extensions to the theory of arrays, such as Wang and Appel's theory of arrays with concatenation \cite{wang_solver_2023}, referred to as \Arrayc, extend the theory of arrays with length, slice, and concat operators. Since this provides arrays with properties similar to those of sequences, mainly 0-indexing and length, we will refer to \Arrayc's arrays as sequences.

\subsection{Known theories}

We present the signatures of \Seqc and \Seqz, starting with the shared symbols of the two, then the specificities of each one:
\begin{itemize}
  \item Common symbols to \Seqc and \Seqz:
        \begin{itemize}
          \item $\sempty$: the empty sequence
          \item $\unit(v)$: a sequence of length $1$ containing only the value $v$
          \item $\slength(s)$: the length of the sequence $s$
          \item $\nth(s, i)$: the value associated with the $i$th index of $s$ if $i$ is within the bounds of $s$, an uninterpreted value otherwise
          \item $\extract(s, i, l)$: the extracted maximal sub-sequence of $s$, starting at $i$ of length $l$ if $i$ is within the bounds of $s$ and $l$ is positive, the empty sequence otherwise
          \item $\sconcat (s_1, ..., s_n)$: the concatenation of the sequences $s_1$, ..., and $s_n$
          \item $\at(s, i)$: a unit sequence containing the $i$th value in $s$ if $i$ is within the bounds of $s$, the empty sequence otherwise
          \item $\contains(s_1, s_2)$: true if $s_1$ is a sub-sequence of $s_2$, false otherwise
          \item $\indexof(s, s', i)$: the first position of $s'$ in $s$ at or after $i$, $-1$ if there are no occurrences
          \item $\replace(s, s_1, s_2)$: the resulting sequence from replacing the first occurrence of $s_1$ with $s_2$ in $s$ if $s_1$ occurs in $s$, $s$ otherwise
          \item $\prefixof(s', s)$: true if $s'$ is a prefix of $s$, false otherwise
          \item $\suffixof(s', s)$: true if $s'$ is a suffix of $s$, false otherwise
        \end{itemize}

  \item Symbols belonging to \Seqc only:
        \begin{itemize}
          \item $\replaceall(s, s_1, s_2)$: the resulting sequence from replacing all occurrences of $s_1$ with $s_2$ in $s$, $s$ if $s_1$ does not occur in $s$
          \item $\rev(s)$: the resulting sequence from reversing $s$
          \item $\update(s_1, i, s_2)$: New sequence of the same size as $s_1$, in which, if $i$ is within the bounds of $s_1$, then the values from $i$ to $i + \slength(s_2) - 1$ are the same values as in $s_2$ and the other values are the same as in $s_1$, otherwise it equals $s_1$
        \end{itemize}

  \item Symbols belonging to \Seqz only:
        \begin{itemize}
          \item $\map(f_n, s)$: the sequence of sort $(\Seq \textsf{ E'})$ resulting from applying $f_n$, which is of sort $(\Array \textsf{ E E'})$ with \textsf{E} as the sort of the values of $s$, to all the values of $s$
          \item $\mapi(f_n, o, s)$: the sequence of sort $(\Seq \textsf{ E'})$ resulting from applying $f_n$, which is of sort $(\Array \textsf{ Int (Array E E')})$ with \textsf{E} as the sort of the values of $s$, to all the values of $s$ and their indices starting from the offset $o$
          \item $\foldleft(f_n, b, s)$: The result of folding over $s$ of sort $\Seq\textsf{ E}$, with an initial value $b$ of sort $\textsf{E'}$, using the function $f_n$ of sort $(\Array \textsf{ E' (Array E E')})$
          \item $\foldlefti(f_n, o, b, s)$: The result of folding over the values of $s$ of sort $\Seq\textsf{ E}$ and their indices, with an initial value $b$ of sort $\textsf{E'}$, using the function $f_n$ of sort $(\Array \textsf{ Int (Array E' (Array E E'))})$, starting from the offset $o$
        \end{itemize}
\end{itemize}

The $\update$ function is described differently in the paper that describes the reasoning implemented in CVC5 \cite{sheng_reasoning_2023} for the theory of sequences and in the documentation of CVC5\footnote{CVC5's sequence theory: \url{https://cvc5.github.io/docs-ci/docs-main/theories/sequences.html}}. In the paper, it is described as a function that sets only the value of one index, and takes that value as a third argument, while in the documentation it takes a sequence as a third argument.

The signature of \Arrayc is the following:

\begin{itemize}
  \item $\lengths_S(s)$: the length of $s$
  \item $\nths_S(i, s)$: the $i$th value of $s$, if $i$ is out of the bounds of $s$ then the value is the default value of the sort of values, the theory assumes that every value sort has a variable corresponding to the default value of that sort
  \item $\repeats_S(v, n)$: a sequence of size $n$ if $n$ is positive, in which all values are $v$, the empty sequence if $n$ is negative
  \item $\apps_S(s_1, s_2)$: concatenates $s_1$ and $s_2$
  \item $\slices_S(i, j, s)$: a sub-sequence of $s$ from $\max(i, 0)$ to $\min(j, l)$, with $l$ as the length of $s$, the empty sequence if such a sub-sequence does not exist
  \item $\mapf_f(s_1, ..., s_k)$: the sequence resulting from applying f element-wise to the $n$ first elements of the sequences $s_1$, ..., $s_k$, where $n$ is $\min(\lengths_S(s_1), ..., \lengths_S(s_k))$
  \item $\updates(i,s,x)$: returns an updated version of $s$ in which $i$ is mapped to $x$ if $i$ is within the bounds of $s$. It is mentioned that the function $\updates$ is reduced to a concatenation of the sequences $\slices(0,i,s)$, $\repeats(v,1)$ and $\slices(i+1,\lengths(s),s)$ when $i$ is within the bounds of $s$.
\end{itemize}

The $\mapf_f$ symbol from \Arrayc is similar to the map function over arrays described in a paper by de Moura and Bjørner presenting the \textsf{CAL} (Combinatory Array Logic) array decision procedure \cite{de_moura_generalized_2009}.

\subsection{Discussion}

The theory of sequences is seen as a generalization of the theory of strings to non-character values. Its signature is largely based on that of the theory of strings. From the available literature \cite{sheng_reasoning_2023} and the properties of sequences in the theory, such as being 0-indexed and having dynamic lengths, we understand that the theory of sequences serves the purpose of more adequately representing arrays as found in programming languages than what the theory of arrays can do. However, a $\sset$ function in \Seq, which would correspond to the $store$ function in the theory of arrays and store one value at one index in a sequence, is missing. This is problematic as that function is usually present in array-like data structures, such as the array assignment operation in C.

Having a rich signature that allows all necessary functions and predicates on values of a theory makes that theory expressive. This means that when users need to express properties in that theory, they will not have to define or axiomatize them if they can rely on the theory's built-in symbols. However, a theory's signature affects how to reason over it. For example, when working with \Seq, it is natural to adapt reasoning over strings to reasoning over sequences, since the signatures of the theory of strings and that of \Seq have many similarities. If only a fragment of \Seq is used, for example, one in which operations over sub-sequences such as $\extract$, $\sconcat$, and $\update$ are not supported, then reasoning over arrays can more easily be adapted to that fragment of \Seq than to \Seq itself. Therefore, having simpler, minimal fragments of a theory can allow for the development of tailored reasoning for that fragment, which can be more efficient than the reasoning used for the whole theory. This is, for example, the case for Integer (resp. Real) Difference Logic, which is a fragment of Integer (resp. Real) Linear Arithmetic and has its own reasoning techniques.

Looking at the \Seqz and \Arrayc theories, the $\textsf{map}$ symbol is present, albeit with different semantics. \Seqz has a $\map$ function and a $\foldleft$ function similar to those commonly encountered in \textsf{ML} and more generally in functional programming languages. The $\map$ function applies a function to all the elements of a sequence and produces a new sequence, while $\foldleft$ iterates over all the elements of a sequence and applies a function to them and to an accumulator that can be of any other sort, returning that accumulator to pass it to the function in the next iterations. Since the SMT-LIB standard version 2.6 does not support higher-order functions, the functions are represented using arrays, and to express complex functions using arrays, it is usually necessary to use quantifiers, which SMT solvers often struggle with. However, such functions would be useful for modeling array-like or list-like data structures from functional programming languages. The $\mapf_f$ function is similar to the one present in the CAL extension of the theory of arrays \cite{de_moura_generalized_2009}, and it applies a function $f$ element-wise to the $k$ elements of $n$ arrays, where $k$ is the length of the smallest of the $n$ arrays, so applying it to one array makes it similar to the $\map$ function from \Seqz.

\section{Design Criteria}

In this section, we will define a set of criteria that we consider important when designing an SMT theory and choosing its signature and semantics.

\subsection{Expressiveness}

Having clear semantics and a rich signature with both the necessary functions to perform commonly encountered operations and the predicates to express the properties that need to be verified is necessary in a theory. It simplifies the work of the users of that theory as they don't have to define or axiomatize additional symbols in the signature of the theory to use it. For example, in the theory of fixed-size bit-vectors, there are bitwise operations like \textsf{bvand} and \textsf{bvor}, as well as bit-vector arithmetic operations like \textsf{bvadd} and \textsf{bvsub}, which are commonly used when working with machine integers.

Written SMT formulae in files, whether by a user or automatically generated by tools like Why3 \cite{filliatre_why3_2013}, Dafny \cite{leino_dafny_2010}, etc., can easily become large and complex. Therefore, expressiveness in theories plays an important role in clarity, understanding, and efficiency. Frequently, when a theory is missing needed features that can be built-in, adding them requires axiomatization with quantifiers, and quantifiers are known to be hard for solvers to work with and can be harder for users to understand.

\subsection{Implementability and efficiency}

The goal of this criterion is to ensure that reasoning over the theory in question is reasonably implementable in an SMT theory combination framework, with the constraints that come with it. Although the implementability and efficiency of a reasoning over a theory depend more on the reasoning itself rather than on the theory, they are significantly affected by the design choices of a theory, as that determines how complex the reasoning would need to be. Having a theory with a concise set of symbols and clear semantics will make it easier to formalize reasoning over it and to implement it. Additionally, theory functions should have well-formalized behavior, with not many special cases in which the behavior significantly changes or becomes more complicated.

This criterion is hard to measure, but when it comes to theory design, it is something that should be taken into account when deciding what signature and semantics to give a theory. It also involves a compromise with expressiveness since making a theory too expressive can complicate the task for those who try to reason over it and implement the reasoning.

\subsection{Avoiding surprises and user-friendliness}

A theory's semantics need to be as clear as possible. The symbols defined in the theory should have straightforward semantics and not many special cases in which they behave differently. Such complexity would make it harder to understand and solve any issues that may be encountered while using the theory. The symbols should also be consistent with one another in the theory, for example regarding associativity (when it is possible to choose) or the order of the arguments. This consistency is evident in the theory of arrays, where both the $\select$ and $\store$ functions take an array term and an index term as the first and second arguments, respectively.

\section{Handling of partial functions}

Partial functions are functions that can be applied outside their domain of definition. To address partial functions, there are three known approaches: underspecification, overspecification, or by adding an argument to be returned when the function is applied outside its domain. In this section, we will clarify these three techniques.

\subsection{Underspecification}

This approach consists of treating values returned by functions when they are applied to arguments for which they are not defined as uninterpreted values. An uninterpreted value is unconstrained and can be any value of the correct sort. Therefore, when checking the satisfiability of an assertion containing such a value, the assertion is either satisfiable, or if it isn't, then it needs to be unsatisfiable for any possible value for that uninterpreted value. This approach was taken for integer and real division. When an integer (resp. real) value is divided by zero in the (Non-)Linear Integer (resp. Real) Arithmetic theory, the resulting value is uninterpreted.

When a user models a program's behavior, the underspecified approach is a safe choice because when a goal is proven to be unsatisfiable, it is proven for any value of the same sort. However, this approach can affect the decidability of a theory. For example, the theory of Real Difference Logic is decidable but becomes undecidable when combined with uninterpreted unary predicates \cite{boigelot_decidability_2023}.

Implementing this approach is not difficult if the framework in which it is implemented has a solver for the theory of uninterpreted functions. Otherwise, dependency on another theory can be problematic for developers of solvers that do not support the theory of uninterpreted functions. Moreover, in the satisfiable case, checking the validity of a model may require the value chosen by the solvers for the uninterpreted value that led to the satisfiable result. Yet, there is no syntax for specifying such values in the SMT-LIB standard version 2.6 \cite{DBLP:conf/smt/BuryB23}.

\subsection{Overspecification}

Another way to handle partially defined functions is by selecting a constant value to be returned when the function is applied outside its domain. For instance, in the theory of Fixed Size Bit-Vectors, the function \textsf{bvudiv} takes two bit-vectors of the same size as arguments and returns a bit-vector of the same size, representing the result of the unsigned division between the two bit-vectors. When the second argument is a bit-vector of zeros, \textsf{bvudiv} returns a bit-vector representing the value $-1$, which contains only ones. The rationale behind this choice is that bit-vectors are commonly used in circuit calculations, and when a circuit receives a zero, it returns all ones\footnote{According to: \href{https://cs.nyu.edu/pipermail/smt-lib/2015/000966.html}{https://cs.nyu.edu/pipermail/smt-lib/2015/000966.html}}. While such justifications apply to some choices of values, it is often unclear which value should be chosen, especially when the chosen value can be obtained by applying the function within its domain.

In the case of the Floating-Point Arithmetic theory, which follows the IEEE standard 754-2008 \cite{noauthor_ieee_2008}, the \textsf{NaN} value is used to represent undefined values. It serves as a catch-all case for the undefined behavior of the theory's functions. However, equivalent values do not exist in other theories. A clear advantage of this approach is that it facilitates the detection of undefined behavior since such cases are not silently treated. Encountering a \textsf{NaN} typically indicates that a partial function was applied outside its domain, although it's important to note that comparing a value to \textsf{NaN} always results in false, so the approach is not entirely foolproof and requires additional checks that values are not \textsf{NaN}.

The primary advantage of this approach is to have a predetermined value as a result whenever the function is applied outside its domain. This provides solvers with less flexibility in how to handle such values, establishing a sort of uniformity in their behavior in such cases. Additionally, it simplifies implementations, as solvers do not need to support the theory of uninterpreted functions or depend on it to handle such cases, they simply need to return the predetermined constant value.

Another example of this approach is seen in the \Arrayc theory with the $\nths_S$ function, which returns a default value when the index is out of bounds of the sequence. The theory assumes that every theory of values has such a default value. However, relying on a default value to be returned can lead to unexpected results. For instance, if applied on integer division, while $\frac{1}{0}=2$ would not be provable, $\frac{1}{0}=\frac{2}{0}$ would be provable since $\frac{1}{0}$ and $\frac{2}{0}$ have the same integer default value.

\subsection{Returned value as an argument}

The idea behind this approach is to let the user decide. Instead of choosing a default value to return or returning an uninterpreted value, the user can choose which value to return by adding an argument to the function, the value of which will be returned when the function is applied outside its domain.

With this approach, if a user needs a partial function to return a specific constant value, they simply need to provide that constant value to the function. If they need an unconstrained value, they can also provide an uninterpreted constant. Implementation-wise, it is straightforward since it involves returning a value provided as an argument whenever the function is applied outside its domain. Moreover, it enhances user-friendliness by giving users control over the returned value when undefined behavior occurs, enabling easy detection and appropriate handling of such cases.

Although not common in the SMT-LIB standard, this approach has been suggested in past discussions on SMT theory design as a preferable solution compared to the previous two approaches, as it represents a compromise between them.

Considering a modified version of the $\nth$ function from \Seq, called $\nth'$, of sort $\Seq -> \Tint -> \Elem -> \Elem$ where this approach is utilized, with the third argument representing the default value. If $\nth'$ is used in a quantification, and underspecification behavior is desired, one way to achieve it is by defining a function $\textsf{nth\_defval}$ of sort $\Seq -> \Tint -> \Elem$ such that $\nth(s, i) = \nth'(s, i,\textsf{nth\_defval(s, i)})$. However, defining $\textsf{nth\_defval}$ is currently not possible in SMT-LIB version 2.6 due to the lack of polymorphism. Therefore, the usage of such an approach may be limited until SMT-LIB version 3 is released, which would allow polymorphism, already supported by some SMT solvers such as Alt-Ergo.

\section{Designing theories of sequences}

In this section, we propose changes to the theory of sequences by considering the criteria outlined in the previous sections. Our version of the theory is based on \Seqc, \Seqz, and \Arrayc.

\subsection{Proposed changes to symbols of the theory of sequences}

Below, we list and explain our modification choices for some sequence functions:

\newcommand\addvmargin{
  \useasboundingbox (-0.5,-0.25) rectangle (2,1.25);
}

\begin{figure}
  \centering
  \begin{tabular}{|p{15mm}|c|c|c|c|} \hline
                                                        &
    no overflow                                         &
    left overflow                                       &
    right overflow                                      &
    left-right overflow
    \\
    \hline \Seqc
                                                        &
    \begin{tikzpicture}[baseline=0]
      \pgfsetcornersarced{\pgfpoint{1mm}{1mm}}
      \draw[fill=white] (0.25,0.5) rectangle (1.25,0.75);
      \draw[fill=black!50] (0,0) rectangle (1.5,0.25);
      \draw[fill=white] (0.25,0) rectangle (1.25,0.25);
      \addvmargin
    \end{tikzpicture} &
    \begin{tikzpicture}[baseline=0]
      \pgfsetcornersarced{\pgfpoint{1mm}{1mm}}
      \draw[fill=white] (-0.5,0.5) rectangle (0.5,0.75);
      \draw[fill=black!50] (0,0) rectangle (1.5,0.25);
      \addvmargin
    \end{tikzpicture}  &
    \begin{tikzpicture}[baseline=0]
      \pgfsetcornersarced{\pgfpoint{1mm}{1mm}}
      \draw[fill=white] (1.0,0.5) rectangle (2.0,0.75);
      \draw[fill=black!50] (0,0) rectangle (1.5,0.25);
      \draw[fill=white] (1.0,0) rectangle (1.5,0.25);
      \addvmargin
    \end{tikzpicture}   &
    \begin{tikzpicture}[baseline=0]
      \pgfsetcornersarced{\pgfpoint{1mm}{1mm}}
      \draw[fill=white] (-0.5,0.5) rectangle (2.0,0.75);
      \draw[fill=black!50] (0,0) rectangle (1.5,0.25);
      \addvmargin
    \end{tikzpicture}
    \\
    \hline Proposal
                                                        &
    \begin{tikzpicture}[baseline=0]
      \pgfsetcornersarced{\pgfpoint{1mm}{1mm}}
      \draw[fill=white] (0.25,0.5) rectangle (1.25,0.75);
      \draw[fill=black!50] (0,0) rectangle (1.5,0.25);
      \draw[fill=white] (0.25,0) rectangle (1.25,0.25);
      \addvmargin
    \end{tikzpicture} &
    \begin{tikzpicture}[baseline=0]
      \pgfsetcornersarced{\pgfpoint{1mm}{1mm}}
      \draw[fill=white] (-0.5,0.5) rectangle (0.5,0.75);
      \draw[fill=black!50] (0,0) rectangle (1.5,0.25);
      \draw[fill=white] (0,0) rectangle (0.5,0.25);
      \addvmargin
    \end{tikzpicture}  &
    \begin{tikzpicture}[baseline=0]
      \pgfsetcornersarced{\pgfpoint{1mm}{1mm}}
      \draw[fill=white] (1.0,0.5) rectangle (2.0,0.75);
      \draw[fill=black!50] (0,0) rectangle (1.5,0.25);
      \draw[fill=white] (1.0,0) rectangle (1.5,0.25);
      \addvmargin
    \end{tikzpicture}   &
    \begin{tikzpicture}[baseline=0]
      \pgfsetcornersarced{\pgfpoint{1mm}{1mm}}
      \draw[fill=white] (-0.5,0.5) rectangle (2.0,0.75);
      \draw[fill=white] (0,0) rectangle (1.5,0.25);
      \addvmargin
    \end{tikzpicture}
    \\ \hline
  \end{tabular}
  \caption{Comparison of the semantics of update in \Seqc and our proposal: the gray sequence is updated using the white sequence at different offsets.}
  \label{fig1}
\end{figure}

\begin{itemize}
  \item $\sget(s, i)$:
        It is similar to the $\textsf{select}$ function of the \Array theory and the $\nth$ (resp. $\nths_S$) function from the \Seqc and \Seqz theories (resp. \Arrayc theory). It returns the value associated with the index $i$ in the sequence $s$ when $i$ is within the bounds of $s$. The function is partial as it is not defined when the index $i$ is out of the bounds of the sequence $s$. In the \Seqc and \Seqz theories, $\nth$ is totalized by underspecification. While in \Arrayc, the default value approach is chosen, by assuming that all value theories have a variable that corresponds to that default value.

        Due to the issues mentioned in the previous section related to the usage of the default argument as a return value for undefined behavior without polymorphism, we chose to follow the underspecification approach by returning an uninterpreted value when $i$ isn't in the bounds of $s$.

        The name of the symbol is $\sget$ and not $\nth$ to make it consistent with the $\sset$ symbol that will be described next.

  \item $\sset(s, i, v)$:
        It is similar to the $\textsf{store}$ \Array theory function and not present in the theories of sequences, but it can be represented in \Seqc with $\update(s, i, \unit(v))$. It is also similar to the $\updates$ function of \Arrayc. It returns a new sequence in which the value $v$ is stored at the $i$th index if $i$ is within the bounds of $s$, and it is undefined otherwise.

        In programming languages, accessing an array-like data structure out of its bounds usually results in an error. It can be argued that the function should show that it failed when it is applied outside its domain. One possible way to represent that failure is by returning the empty sequence. However, the semantics of the theory do not necessarily need to exactly follow the semantics of the programming languages they are used to prove the soundness of. Additionally, to use axioms from the \Array theory decision procedures to reason over sequences, such as the \textsf{select-over-store} \cite{christ_weakly_2015,de_moura_generalized_2009} which dictates:
        \[
          \inference[\textsf{select-over-store}] { a = store(b, i, v) & w = select(a, j)  & a,b: \Array, i, j: \Tint, v, w: \Elem } {
            (i = j \land v = w) \lor (i \neq j \land select(a, j) = select(b, j))
          }
        \]

        A similar axiom in the theory of sequences would state that setting the value of an index only changes the value associated with that index if it is within the bounds of the sequence and does not affect the other indices. Therefore, to support such axioms, returning $s$ when $i$ is out of the bounds of $i$ seems like a better choice, and it is the chosen approach for this function. $\sset$ is then defined by the following axiom:
        \begin{gather*}
          s_2 = \sset(s_1,i,v) \equiv \\
          \slength(s_2) = \slength(s_1) \land\\
          \forall j:\Tint, 0 \le j < \slength(s_1) \implies \sget(s_2,j) = \ite(j = i, v, \sget(s_1, j))
        \end{gather*}

  \item $\sconst(l, v)$: equivalent to $\repeats$ in \Arrayc, it is defined by the following axiom:
        \begin{gather*}
          s = \sconst(l,v) \equiv \\
          \ite (l \le 0, s = \sempty, \\
          \slength(s) = l \land \forall i:\Tint, 0 \le i < l \implies \sget(s,i) = v)
        \end{gather*}

  \item $\slice(s, i, j)$:
        Similarly to the $\extract$ function from the \Seq, \Seqc, and \Seqz theories and the $\slices_S$ function in the \Arrayc theory. The purpose of this function is to make it possible to extract a subsequence from a sequence. It takes the target sequence argument $s$ and two Integer arguments $i$ and $j$. In the \Seqc and \Seqz theories, the choice was made for $i$ to be the first index of the subsequence and $j$ to be the length of the subsequence, likely to stay consistent with the substring extraction $\textsf{str.substr}$ in the string theory. While the $\slices_S$ returns the subsequence from index $i$ to index $j - 1$.

        Each one of the two versions, first index and length, and first index and last index, can be expressed by the other one. It is unclear if one is better than the other. An advantage of having the length as an argument is that it is not needed to compute it to set the length constraint. On the other hand, if the reasoning can more easily get the slice of the sequence using the first and last index, then it might be better to have them as arguments instead of having to compute the last index.

        In our case, we prefer a variation of the second version, one that is similar to the $\textsf{extract}$ function from the Fixed-Size Bitvector theory and extracts the subsequence from the index $i$ to the index $j$, because it seems more natural to reason over a slice of a sequence from its first index to its last index than with its first index and its length, but that is pretty arbitrary and as said before, they both can express one another.

        Given the first and last indices $i$ and $j$, the $\slice$ function is partial since it is defined only when $0 \le i \le j < \slength(s)$. That gives it four special cases to deal with:
        \begin{itemize}
          \item The negative length slice case, when $j < i$
          \item The left overflow case, when $i < 0$
          \item The right overflow case, when $\slength(s) \le j$
          \item The left-right overflow case, when both $i < 0$ and $\slength(s) \le j$ are true
        \end{itemize}

        In the case of the negative length slice, the natural solution would be to return an empty sequence since we are effectively trying to get a slice of non-strictly positive length. For the overflow cases, the suggested solution in the \Arrayc theory seems best as it introduces consistency in how they are handled. It consists in taking the maximal subsequence of $s$ that can be obtained between $i$ and $j$, by selecting as the first index of the resulting slice $\max(i, 0)$ and as the last index $\min(j, \slength(s) - 1)$. By following that choice, the resulting axiom of the function is:
        \begin{gather*}
          s_2 = \slice(s_1,i,j) \equiv \\
          \ite (i \le j,
          \llet\, i' = \max(i,0), \llet\, j' = \min(j,\slength(s_1) - 1),
          \\\slength(s_2) = j' - i' + 1 \land \forall k: \Tint, i' \le k \le j' \to \sget(s_2, k - i') = \sget(s_1, k),
          \\s_2 = \sempty)
        \end{gather*}

  \item $\update(s_1, i, s_2)$:
        Like the $\update$ function from \Seqc, which we will refer to as $\update_{\textsf{cvc5}}$. It updates a sequence $s_1$ starting from the index $i$ with the sequence $s_2$. It is only defined when $0 \le i < i + \slength(s_2) \le \slength(s_1)$ and has similar special cases to the $\slice$ function:
        \begin{itemize}
          \item The empty sequence case when $\slength(s_2) = 0$
          \item The left overflow case when $i < 0 \le i + \slength(s_2)$
          \item The right overflow case when $\slength(s_1) \le i + \slength(s_2)$
          \item The left-right overflow case when both overflow conditions are true
        \end{itemize}

        The choice of the \Seqc theory is to have $\update_{\textsf{cvc5}}$ behave as an iteration of $\sset$ on $s_1$ that goes from $i$ to $i + \slength(s_2) - 1$, writing the values of $s_2$, and the iteration stops when $i$ is not in the bounds of $s_1$. So in the case of left overflow, the returned value is $s_1$, in the case of right overflow, an intersection is done between $s_1$ and $s_2$ if $i < \slength(s_1)$; otherwise, $s_1$ is returned. $s_1$ is also returned when $s_2$ is empty. The $\update_{\textsf{cvc5}}$ function's axiom is:
        \begin{gather*}
          s = update_{\textsf{cvc5}}(s_1, i, s_2) \equiv \\
          \slength(s) = \slength(s_1)\, \land \\
          \ite(0 \le i < \slength(s_1), \\
          \forall j: Int, 0 \le j < \slength(s_1) \implies \\
          \sget(s, j) =  \ite(i \le j < i + \slength(s_2), \sget(s_2, j - i), \sget(s_1, j)), \\
          s = s_1)
        \end{gather*}

        In our case, we see $\update$ more like a bag of $\sset$ operations, and we think that it would be simpler for it to respect the following axiom, which removes the condition that $i$ needs to be in the bounds of $s_1$:
        \begin{gather*}
          s = \update(s_1, i, s_2) \equiv \\
          \slength(s) = \slength(s_1)\, \land\\
          \forall j: Int, 0 \le j < \slength(s_1) \implies  \\
          \sget(s, j) = \ite(i \le j < i + \slength(s_2), \sget(s_2, j - i), \sget(s_1, j))
        \end{gather*}

        In addition, to follow the least surprise criteria, this also makes it behave consistently with $\slice$ on how the left, right, and left-right overflow cases are treated, which is by taking the intersection between $s_1$ and $s_2$ offset to the $i$th index. Figure \ref{fig1} illustrates the difference.

  \item $\map(f, s_1, ..., s_n)$: The semantics for $\map$ are the same as the semantics of the $\mapf_f$ in \Arrayc, as it is an n-ary version of the $\map$ function in \Seqz:
        \begin{gather*}
          s = \map(f, s_1, ..., s_n) \equiv \\
          \slength(s) = k \land \\
          \sget(s, 0) = f(\sget(s_1, 0), ..., \sget(s_n, 0)) \land \ldots \land\\
          \sget(s, k-1) = f(\sget(s_1, k-1), ..., \sget(s_n, k-1)) \\
        \end{gather*}

        Where $k$ is the length of the shortest of the sequences $s_1$ to $s_n$.

  \item $\mapi(f, o, s_1, ..., s_n)$: A version of $\map$ in which the mapped function is applied starting from an offset $o$:
        \begin{gather*}
          s = \mapi(f, o, s_1, ..., s_n) \equiv \\
          \ite(o \ge k, s = \sempty, \\
          \slength(s) = k - o \land \\
          \sget(s, 0) = f(o, \sget(s_1, o), ..., \sget(s_n, o)) \land \ldots \land\\
          \sget(s, k - o) = f(k-1, \sget(s_1, k-1), ..., \sget(s_n, k-1))) \\
        \end{gather*}
        Where $k$ is the length of the shortest of the sequences $s_1$ to $s_n$. In fact, it can be reduced to $\map(f, s_o, s_1, ..., s_n)$ where $s_o$ is a sequence of integers of size $k - o$, containing values going from $o$ to $k-1$, when $k$ is greater than $o$ and $o$ is positive.

\end{itemize}

\subsection{Theory of sequences proposal}

As to not stray too far from what already exists, our proposal consists in combining the aforementioned theories: \Seqc, \Seqz and \Arrayc, and applying our proposed changes. We outline our proposal in Figure \ref{fig:seqtheory}.

\begin{figure}[!htbp]
  \centering
  \noindent\begin{tabular}{  p{0.13\textwidth} | p{0.37\textwidth} | p{0.4\textwidth}}
    \hline
    Symbol        & Sort
                  & Remark
    \\ \hline
    $\sempty$     & $\Seq$
                  &
    \\
    $\sconst^{*}$ & $\Tint -> \Elem -> \Seq$
                  & replaces $repeats_S$(\Arrayc)
    \\
    $\unit$       & $\Elem -> \Seq$
                  & $\unit(v) = \sconst(1,v)$

    \\
    $\slength$    & $\Seq -> \Tint$
                  &
    \\
    $\sget^{*}$   & $\Seq -> \Tint -> \Elem$
                  & replaces $\nth$ (\Seqc and \Seqz) and $\nths_S$ (\Arrayc)
    \\
    $\sset^{*}$   & $\Seq -> \Tint -> \Elem -> \Seq$
                  &
    \\
    $\slice$      & $\Seq -> \Tint -> \Tint -> \Seq$
                  & replaces $\extract$ (\Seqc and \Seqz) and $\slices_S$ (\Arrayc)
    \\
    $\sconcat$    & $\Seq -> \Seq -> \Seq$
                  & replaces $\sconcat$ (\Seqc and \Seqz) and $\apps_S$ (\Arrayc)
    \\
    $\at$         & $\Seq -> \Tint  -> \Seq$
                  &
    \\
    $\contains$   & $\Seq -> \Seq -> \Tbool$
                  &
    \\
    $\replace$    & $\Seq -> \Seq -> \Seq$
                  &
    \\
    $\indexof$    & $\Seq -> \Seq -> \Tint$
                  &
    \\
    $\prefixof$   & $\Seq -> \Seq -> \Tbool$
                  &
    \\
    $\suffixof$   & $\Seq -> \Seq -> \Tbool$
                  &
    \\
    $\replaceall$ & $\Seq -> \Seq -> \Seq$
                  & Only present in \Seqc
    \\
    $\rev$        & $\Seq -> \Seq$
                  & Only present in \Seqc
    \\
    $\update^{*}$ & $\Seq -> \Tint -> \Seq -> \Seq$
                  & replaces $\update$ (\Seqc)
    \\
    $\map^{*}$    & $(\Elem_1 -> ... ->  \Elem_n -> \Elem') -> \Seq_1 -> ... ->  \Seq_n -> \Seq'$
                  & replaces $\mapf_f$ (\Arrayc), and $\map$ (\Seqz)
    \\
    $\mapi^{*}$   & $(\Tint -> \Elem_1 -> ... ->  \Elem_n -> \Elem') -> \Tint -> \Seq_1 -> ... ->  \Seq_n -> \Seq'$
                  & replaces $\mapi$ (\Seqz)
    \\
    $\foldleft$   & $(\alpha -> \Elem -> \alpha) -> \alpha -> \Seq -> \alpha$
                  & Only present (\Seqz)
    \\
    $\foldlefti$  & $(\Tint -> \alpha -> \Elem -> \alpha) -> \Tint -> \alpha -> \Seq -> \alpha$
                  & Only present (\Seqz)
    \\

    \hline
  \end{tabular}
  \caption{Signature of the proposed Sequence theory. The $^{*}$ after a symbol means that the symbol's semantics are as described in the previous subsection. The sequence sorts $\Seq_1 ... \Seq_n$ and $\Seq'$ have elements of the sorts $\Elem_1 ... \Elem_n$ and $\Elem'$ respectively.}
  \label{fig:seqtheory}
\end{figure}

\subsection{Fragmenting the theory of sequences}

As demonstrated in Wang and Appel's decision procedure \cite{wang_solver_2023}, which was used to reason over arrays in C programs, and in the calculus developed by Sheng et al. \cite{sheng_reasoning_2023}, which was used to reason over vectors from smart contract verification. In some cases, a smaller fragment of the theory of sequences is sufficient to reason over arrays from many programming languages. Defining such a fragment is crucial, especially for solver developers, as it allows them to focus on developing reasoning capabilities for only the subset of the theory they require.

Given the considerable variation in operations supported by array-like data structures in programming languages, determining the precise set of symbols necessary and sufficient to represent such structures is challenging.

Building upon \Arrayc's extension of the theory of arrays, which has been proven decidable when the verification condition has no index shifting\footnote{Terms of the form: $\forall i: \Tint, 0 \le i < \lengths_S(s) - n \implies \nths_S(i, s) = \nths_S(i + n, s)$, where $s$ is a sequence and $n$ an integer literal} \cite{wang_solver_2023}, we can define a fragment of the theory of sequences based on our proposal, which can be reducible to the symbols of \Arrayc. The fragment is presented in Figure \ref{fig:seqtheoryfragment}.

\begin{figure}[!htbp]
  \centering
  \noindent\begin{tabular}{  p{0.25\textwidth} | p{0.65\textwidth} }
    \hline
    Symbol                   & Reduction
    \\ \hline
    $\sempty$                & $\repeats(\_,0)$
    \\
    $\sconst(l, v)$          & $\repeats(v, l)$
    \\
    $\unit(v)$               & $\repeats(v,1)$
    \\
    $\slength(s)$            & $\lengths(s)$
    \\
    $\at(s,i)$               & $ite(\nths(s,i) = \delta, \repeats(\_,0), \repeats(\nths(s,i),1))$
    \\
    $\sget(s,i)$             & $\nths(s,i)$
    \\
    $\sset(s,i,v)$           & $\updates(i,s,v)$
    \\
    $\slice(s,i,j)$          & $\slices(s,i,j)$
    \\
    $\sconcat(s_1,s_2)$      & $\apps(s_1,s_2)$
    \\
    $\update(s_1, i, s_2)$   & $\apps(\slices(s_1, 0, i\mathord{-}1),$ \linebreak$\apps(s_2,\slices(s_1,i+\slength(s_2), \slength(s_1)-1)))$
    \\
    $\map(f, s_1, ..., s_2)$ & $\mapf_f(s_1, ..., s_2)$
    \\
    \hline
  \end{tabular}
  \caption{Fragment of the proposed theory of sequences and how its symbols can be reduced to those of \Arrayc. $\delta$ represents a default value of the same sort as the return value of $\nths$, which is returned when $\nths$ is applied outside its domain. $\_$ any value of the right sort.}
  \label{fig:seqtheoryfragment}
\end{figure}

Other symbols can also be added, such as $\mapi$, which, as mentioned previously, can be reduced to $map$. Additionally, symbols like $\foldleft$, $\foldleft$, and $\rev$ are common in array-like data structures, especially in functional programming languages. While such functions can always be defined as recursive functions, the potential impact of adding them to this fragment on decidability and reasoning efficiency needs to be further explored.

A proof of the soundness of the reductions of the symbols in the fragment is necessary, but we consider this to be beyond the scope of our paper. Nevertheless, this example can serve as a step towards defining a fragment of the theory of sequences tailored to represent array-like data structures.

\section{Conclusion}

In this paper, we have explored SMT theory design through the lens of the theory of sequences. We defined a set of criteria to consider when designing a theory and examined how partial functions are handled. Additionally, we proposed a variant of the theory of sequences based on these criteria, which we believe should be considered in the event of the theory's standardization.

While the criteria we outlined should aid in theory design, they are not precise enough and are open to interpretation, particularly regarding notions such as user-friendliness. Ultimately, choices in theory design can be quite subjective.

\newpage

\bibliographystyle{plain}
\bibliography{bib}

\end{document}